# Exploitation All the Way Down: Calling out the Root Cause of Bad Online Experiences for Users of the "Majority World."[1]

Hellina Hailu Nigatu, UC Berkeley, USA[2]
Zeerak Talat, Edinburgh University, UK

**Abstract**
Global Majority users are exposed to multitudes of harm when interacting with online platforms. This essay illuminates how exploitation in the advances of Artificial Intelligence is tied to historical exploitation and how the use of blanket terminology overshadows the layers of exploitation and harm "Global Majority" populations face. It first discusses the multitude of harm content moderators from the Global Majority face, arguing against the current trend of protection through exploitation, then it illustrates the nuances and differences within the Global Majority, and finally, it outlines actionable items to move away from such harm.

# Introduction

Global Majority users are disproportionately affected by the more extreme harms caused due to harmful content online. For instance, failures in moderation on Facebook have resulted in physical harm and escalation of violence in countries like Myanmar and Ethiopia (Akinwotu, 2021) the spread of misinformation on WhatsApp led to violent attacks on minorities in India (Samuels, E. 2020); and YouTube users from countries that do not have English as their primary language are at 60% higher rate of being exposed to content they will "regret" watching (McCrosky et. al. 2021). Such lackluster moderation and failure of automatic detection for the majority of the world's languages emboldens malicious content creators to post policy-violating videos (Nigatu et. al, 2024).

Platforms use a combination of automated systems and human moderators to moderate content (Roberts 2019). Generally, automated content moderation involves using trained machine learning models to determine if a post should be sanctioned due to breaches of policy, e.g., on hate speech and toxicity. However, not all users are protected equally (Dias Oliva, 2020). The field of natural language processing (NLP) has paid little attention to non-European languages, which has lead to a lack of data and technological resources to train robust automated detection systems. Moreover, platforms focus their efforts disproportionately on Western countries. For instance, in 2020 while 90% of its users live outside of the United States (US) and Canada, Meta (then Facebook) spent 87% of its time moderating posts in the US (Tworek, 2021). Such disparity is also reflected in moderation personnel: YouTube reports that 89.2% of its human moderators operate in English (Google, 2023), neglecting that 67% of videos are posted exclusively in languages other than English and 5% in multiple languages including English (Van Kessel et al, 2019).

---

[1] Published as a chapter in Official Outcome of the UN IGF Data and Artificial Intelligence Governance Coalition report on AI from the Global Majoirty:
https://www.intgovforum.org/en/filedepot_download/279/28447
[2] Corresponding author hellina_nigatu@berkeley.edu

The harm that speakers of the majority of the world's languages face in relation to content moderation extends beyond exposure to harmful content as users of online platforms. Big Tech companies hire content moderators from the Global Majority, which appears like an increased effort to protect users from those communities. However, these moderators often operate under deplorable working conditions and without fair compensation for conducting deeply traumatizing work (Perrigo, 2022). Such workers, who are often employed from African, South American, South East Asian, and South Asian countries, also provide labeled data for guardrails of Large Language Models like ChatGPT (Perrigo, 2023), models which do not work well in languages spoken by the Global Majority (Ojo et al., 2023), or are entirely unavailable.

Understanding and implementing effective policy to protect users of Global Majority must begin by uncovering what lies beneath blanket terminology that serves to obscure nuances; starting with the term Global Majority. While the term has been adopted as a reclaiming of power by appealing to the number of people grouped under it, it is still a blanket term covering several geographies, hundreds of cultures, and thousands of languages whose common predicament is exploitation by the powers on the other side – a concern that remains unresolved by the adoption of the term. Prior work has demonstrated the cultural nuances that result in the under-moderation or over-moderation of online users from the "Global Majority" or "Global South" (Shahid et al, 2023). Hence, to effectively impact practical policies, we must start by examining these nuances and uncovering what is underneath the blanket terminologies.

In this essay, we first dive deeper into multitudes of harm faced by content moderators from the Majority world, reflecting on how the common denominator is exploitation. Then, we examine the current alternatives in online moderation which pose a false dichotomy for moderation to be effective, for which surveillance is an inevitable consequence. We call out the root problem that presents these alternatives as the only options. Next, we detail the social, political, and economic structures within the "Global Majority" to illustrate the nuances in different communities that would render blanket policies ineffective. Finally, we put forth a call to action to ensure the effective protection of "Global Majority" users on online platforms. We argue that what ties the experience of Global Majority people is the continued exploitation and disregard for well-being by Big Tech and states outside of the Global Majority, which bears similarities to exploitation by colonial bodies during the period of European colonization.

# Discussion

## The Cycle of Harm In Moderation and Inclusion

In 2021, Meta (then Facebook) faced scrutiny after a whistleblower, Frances Haugen, leaked internal documents detailing the harms the platform was fostering, in some cases not taking action to rectify the situation even after becoming aware of it (Horwitz, 2021). One trend in the moderation landscape has been to hire moderators in Global Majority countries, sometimes through third-party companies. However, the working conditions of the moderators are usually dire (Perrigo, 2022). While cases brought directly against companies like Microsoft and Meta have resulted in settlement payments and some policy changes for moderators hired directly by the companies (Newton, 2020), moderators hired by third-party companies risk mass layoffs and threats against forming unions (Perrigo, 2022). This double standard is a parallel to other exploitative work performed in "Global Majority" countries (e.g. the externalization of "Global Minority" pollution and trash to the "Global Majority" (Liboiron, 2021)), where workers are treated differently for the same work when it is performed in "Global Minority" countries. The exploitation does not stop there. Perhaps ironically, such moderators are hired to moderate OpenAI models like ChatGPT, which do not work for the African languages that they speak (Ojo et al, 2023). In fact, ChatGPT was not available in countries like Ethiopia until November 2023 (Shega, 2023). In this way, the labor of the "Global Majority" is extractive, and the conditions under which moderators work are for the benefit of the privileged few who can operate the internet in languages like English and Spanish.

Communities from the "Global Majority" are exposed to harm (1) while using the platforms, due to weak platform policy enforcement and limited performance of technologies used in the moderation pipeline; (2) while moderating harmful content by virtue of exposure to traumatic content; (3) through poor working conditions and exploited labor; and (4) through technologies that exploit their labor but leave out their whole communities from whatever benefit the technology might provide. At the center of this cycle of harm is the exploitation and neglect of the wide swath of communities. The current systems that sustain the digital landscape are an extension of the history of colonization and exploitation that have ravaged the "Global Majority" (Kwet, 2019). Even when these communities are included in Artificial Intelligence research, they are treated as "bottom billion petri dishes"(Sambasivan et al, 2021, p.320)–their diversity and the weak policies protecting them make them an attractive test-bed for evaluating model robustness with little-to-no consequence or cost.

## False Dichotomies of Harm: Either you are surveilled or you are left in the trenches.

Communities that have largely been excluded from policy and technological advances in the moderation space are exposed to harmful content daily. These unmoderated harmful content could be due to (1) policies that exist but are not enforced properly for these communities, or (2) policies that do not exist since the design of policies takes place under contexts that do not account for the diverse realities of "Global Majority." When policies do exist and are under-enforced, malicious actors exploit the under-enforcement to propagate policy-violating content. As such, communities who have already been exploited by global structures are exploited again in our failure to effectively moderate online spaces.

When policies that reflect the diverse cultural context in the "Global Majority" simply do not exist, entire communities and cultures are left in a vacuum. Indeed, some companies seek to enforce a single standard upon all users, disregarding cultures, customs, and traditions. For instance, Facebook's one-size-fits-all approach resulted in the removal of a post of village kids swimming in a pond for violating the platform's policy against child nudity; although in the context of the poster, it is a common activity for children to swim naked in their local ponds to avoid "being scolded by their parents" (Shahid & Vashistha, 2023, p. 5).

With the rapid advances of Large Language Models and the "low-resource language" NLP community trying to increase the representation of these languages, harmful, toxic, and culturally nonrepresentative content on online spaces risks trickling down to model development and deployment. Generative models are trained using data from YouTube, Twitter, and general web scraping (Cole, 2024). However, training models for the majority of the world's languages present a particular risk as effective content moderation technologies and practices are not deployed for such languages. Thus, risks of harm are compounded by a lack of appropriate moderation, thereby compounding the risks of harm that have been documented for English (Talat et al. 2022).

Platforms that benefit from their users should adhere to their end of the bargain and provide a "positive experience for everyone on [their] platforms no matter where they [the users] are in the world" (Google, 2023, p. 8). Effective content moderation infrastructures, both human and automated, are required for safely building language technologies and content moderation technologies. However, many language technologies have risks of dual-use (Kaffee et al., 2023), including the risk of surveillance (Solaiman et al. 2023). It is therefore particularly important to consider how technologies are deployed and used, in addition to how data is gathered for the technologies themselves.

Here we would like to pause and reflect on what exactly effective moderation is, especially in the current context of the moderation pipeline. If the premise of moderation was not capitalistic and exploitative, could we have safer online experiences that put the power in users and not in companies that are out for profit?

## What Lies Under Blanket Terminologies?

The degree and type of harm communities from the Global Majority face are shaped by the social, political, and economic realities of each community. Take two YouTube users studied by Nigatu & Raji, (2024) who studied the experiences of Ethiopian women on YouTube: a migrant domestic worker and a software engineer in the United States. Both users are Ethiopians, women, and of the Global Majority; yet have completely different realities. Migrant domestic workers cross borders to countries like Qatar and Lebanon en masse, either legally or via human traffickers. Once there, most of these women are subject to inhumane treatment, and sexual harassment and are often left without access to legal or medical services (Diab et al., 2023). Nigatu & Raji, (2024) show how these migrant domestic workers are exposed to harm through exposure to graphic and sexual videos while seeking medical help on online platforms. On the other hand, the Ethiopian Software Engineer living in the US is exposed to the same policy-violating content as the migrant workers when they search in their language. That is, a shift of location does not indicate a shift in types of policy-violating content. Change in policy enforcement might, for instance, remove policy-violating posts that expose both sets of users to harm. However, removal would not satisfy the need for information from the migrant worker, in this case, medical advice.

Political responses of different countries towards platform policies, or failures of platform policies also vary drastically. Countries like Ethiopia, Somalia, and Sudan ban online platforms when policies do not align with their values or when policies do not protect citizens from violent content. However, this has little impact on the actual problem as users resort to VPN services to access the platforms. Additionally, representatives for these platforms are most often subject to regulatory scrutiny in Global Minority countries, even when the harms are primarily impacting people in the Global Majority. It is clear the platforms respond to the callouts by powerful governments; Europe has constantly been praised for the GDPR and its requirements against online harm to its citizens.

While the term "Global Majority" is an evolution from prior binaries based on social and economic status or geographic location (Khan et al., 2022), it is still a binary. The realities–and needs–of Indigenous and Aboriginal communities who continue to suffer the consequences of colonization and occupied land are different from those of African and Asian countries that faced the brunt of exploitation colonialism. Within the Global Majority several layers of class, ethnicity, and power result in the exploitation and harm of some communities over others. There is no single "AI from the Global Majority" because the "Global Majority" is many.

**Call to Action:** Throughout this essay, we have discussed the degree and depth of harm and exploitation that Global Majority users face. However, Global Majority users are not idly waiting for the mercy of the powers that be; to the extent that they can, they devise ways to protect themselves from harm[3]. We can augment their efforts by designing interventions that support them and relying on methods like participatory design as we build AI tools. Additionally, members of the Global Majority face layers of barriers to entering academic and policy spaces at a Global scale (Septiandri et al., 2023). Those who do make it, ourselves included, have degrees of privilege not afforded to the many who are organizing on the ground. Hence, we are responsible for engaging with community organizations–to the degree they are interested–to connect the academic and policy space with community organizing.

---

[3] Instagram users create Fake-Instagram or "Finsta" accounts to share more intimate content with a close group of friends. A YouTube user in Nigatu & Raji (2024) study created multiple accounts for different aspects (religious, educational, and general) becuase she did not "want to be hit with disturbing content when [I] was watching a religious sermon or looking at a lecture."

## Conclusion

The manifold of communities that the "Global Majority" encompasses makes it challenging to enforce one-size-fits-all policies. The harms members of these communities face vary across the diverse social, economic, and political axes each community has. Most of the current policies for protecting users in the digital age have been designed, tried, and tested in the "Global Minority" context. Our response to the fact that we have ignored the majority of the world's population in policy making and implementation should not be to blindly extend these policies to the communities we ignored. In moving from neglect to blind inclusion, we risk the exploitation of community members at several levels of the pipeline. Instead, we should focus our efforts on augmenting community efforts and building interventions that center community needs.

**Notes**